\documentclass[a4paper]{article}

\usepackage{amsmath,amstext,amsgen,amsbsy,amsopn,amsfonts,amssymb}
\usepackage{array}
\usepackage{easybmat}
\usepackage{graphics}
\usepackage[pdftex]{graphicx}
\usepackage{epsfig}
\usepackage{amsthm}
\usepackage{bm}
\usepackage{algorithm}
\usepackage{algorithmic}

\usepackage{wrapfig}
\usepackage{hyperref}

\begin{document}
\large
\title{\textbf{A First Course in Data Science}}
\author{    Donghui Yan and Gary E. Davis\\[0.07in]
            Department of Mathematics and Program in Data Science\\
            University of Massachusetts Dartmouth
        }

\date{\today}
\maketitle 
\normalsize
\begin{abstract}
\noindent {\it Data science} is a discipline that provides principles, methodology and guidelines for the analysis of data for tools, values, or insights. Driven by a huge workforce demand, many academic institutions have started to offer degrees
in data science, with many at the graduate, and a few at the undergraduate level. Curricula may differ at different institutions, because of varying levels of faculty expertise, and different disciplines (such as Math, computer science, and business etc) in developing the curriculum. 
The University of Massachusetts Dartmouth started offering degree programs in data science from Fall 2015, at both the undergraduate and the graduate level. Quite a few articles have been published that deal with graduate data science courses, much less so dealing with undergraduate ones. Our discussion will focus on undergraduate course structure and function, and specifically, a first course in data science. Our design of this course centers around a concept called the {\it data science life cycle}. That is, we view tasks or steps in the practice of data science as forming a process, consisting of states that indicate how it comes into life, how different tasks in data science depend on or interact with others until the birth of a data product or the reach of a conclusion. Naturally, different pieces of the data science life cycle then form individual parts of the course. Details of each piece are filled up by concepts, techniques, or skills that are popular in industry. Consequently, the design of our course is both ``principled" and practical. A significant feature of our course philosophy is that, in line with activity theory, the course is based on the use of tools to transform real data in order to answer strongly motivated questions related to the data.
\end{abstract}
\section{Introduction}
\label{section:Intro}
We discuss our implementation of a first-year undergraduate course in data science as part of a 4-year university-level BS in data science, and we also elaborate what we see as important principles for any beginning undergraduate course in data science. Our principal aim is to stimulate discussion on relevant principles and criteria for a productive introduction to data science.
\section{Background on data science}
The term ``data science" was coined by Jeff C. Wu in his Carver Professorship lecture at the University of Michigan in 1997 (Wu 1997). In this and a subsequent 1998 Mahalanobis Memorial Lecture (Wu 1998), Wu advocated the use of data science as a modern name for statistics. This is the first time the term ``data science'' was used in the statistical community. Cleveland (Cleveland 2001) outlined a plan for a ``new'' discipline, broader than statistics, that he called ``data science'', but did not reference Wu's use of the term. The International Council for Science: Committee on Data for Science and Technology began publication of the {\it Data Science Journal} in 2002, and  Columbia University began publication of {\it The Journal of Data Science} in 2003.
\\
\\
Data science became popular during the last decade with the booming of many major Internet corporations, such as {\it Yahoo, Google, Linkedin, Facebook} and {\it Amazon}, and many start-ups built from data, such as {\it Palantir, Everstring, the Climate Corporation}, and {\it Stitch Fix}. Nowadays, ``data science", along with ``big data", has become one of the most frequently used phrases in venues such as business, news, media, social networks, and academia,
with ``data scientist'' becoming one of the most popular job titles (Davenport and Patil 2012, Columbus 2018). 
\\
\\
Despite the fact that data science has become so popular, and we are using products enabled by data science on almost a daily basis, there is currently no consensus on the definition of data science. While Wu's proposal of the use of the name ``data science'' adds a modern flavor to traditional statistics, we, along with a majority of working data scientists, consider data science as a broader concept than statistics. We view {\it data science} as the science of learning from data: a discipline that provides theory, methodology, principles, and guidelines for the analysis of data for tools, values, or insights. Here tools may include those that can help the user for better analysis, such as tools for visualization, data collection or exploration, and value refers mainly to those with commercial or scientific value. 
\\
\\
Our view of data science has ingredients from several sources, including traditional statistics---Leo Breiman's {\it ``two cultures''} argument of modeling (Breiman 2001)---and, in terms of coverage of topics, David Donoho's {\it ``50 years of data science"} lecture at Princeton University, 2015 (Donoho 2015; see also Donoho 2017). In particular, our view of data science consists of both the generative and the predictive ``culture''. Effectively, this would include machine learning---mostly with a predictive nature---as part of data science, thus putting these two subjects of learning from data, namely, statistics and machine learning, under a common umbrella. This allows a unified treatment of a wide range of problems, including estimation, regression, classification, ranking, as well as unsupervised (or semi-supervised) learning under the broad term ``modeling'' (or analysis). The benefit is immediate: developments and expertise in these two historically separate subjects could inform each other, and many redundant course offerings due to administrative barriers can be removed. Another crucial element in our view is that, one could start with a large amount of data without any particular questions in mind, and relevant questions would be figured out while exploring the data. 
This is what drives the recent surge of interests in data science, given the prevalence of data generating sources such as the Internet, mobile and portable devices, and the increasing feasibility of collecting large amounts of data. A third point is that, data science should also include an interface layer that interacts with domain knowledge or the business aspect, and also algorithms or techniques that deals with the implementation, that is, the computer science aspect. So, in our view, data science is an interdisciplinary subject that encompasses the traditional regimes of statistics and machine learning, business or domain sciences, and computer science.
\section{Introductory undergraduate data science courses}
Driven by a huge demand in data science (Manyika et al., 2011; PwC; Columbus, 2017), many academic institutes have started offering degrees in data science, with many at the graduate and a few at the undergraduate level (see, for example, National Academies 2018). The curriculum may differ at every institute, due possibly to the fact that there is still no consensus in the definition of data science. 
At the University of Massachusetts, Dartmouth, we started offering a BS and MS in data science from Fall 2015. Quite a few articles have been published that discuss data science courses, e.g., Tishkovskaya and Lancaster (2012), Baumer (2015), Escobedo-Land and Kim (2015), Hardin et al. (2015), and Horton et al. (2015). Our discussion here will be about undergraduate data science and, more specifically, a first course in data science (labelled as ``DSC101'' at the University of Massachusetts Dartmouth). Such a course gives an overview and brief introduction to the concepts and practices of data science, and serves three goals. 
\begin{itemize}
\item It introduces to students the notion that data entails value, thus helping motivate students to the study of data science. 
\item It provides students with a big picture and basic concepts of data science, as well as the main ingredients of data science. 
\item Students will learn some practical techniques and tools that they can apply later in more advanced courses or when they start work after their degree program. 
\end{itemize}
Our curriculum design centers around the {\it data science life cycle} and is not simply a loose collection of various topics in data science. It is based on a process model. The idea is that we view individual steps or tasks in data science as forming a process where some may depend on, or interact with others, or may repeat as more insights are gained along the way, until the reach of a conclusion or the birth of a data product\footnote{A data product is any product built from the data. It can be a piece of software (such as a recommendation
system in an e-commerce web), a collection of data that some vendors use to make profit (for example, personal data processed
from data crawled from many different sources and arranged in tabular format, such as \url{https://www.truthfinder.com}), or a 
software tool that one can use to carry out the analysis for a specific application. }. A brief introduction to each piece in the process then forms the individual parts of DSC101, with details to be covered in more specialized or advanced courses. The design of a data science course could also be based on case studies. There are courses in statistics designed with this approach, for example, Nolan and Speed (2000). However, we have not seen many data science courses designed this way; the exceptions are Hardin et al. (2015) and Nolan and Temple Lang (2015). A case study based approach would require a careful selection of study cases with each emphasizing a different aspect of data science so as to ensure coverage of the course on data science topics, which is far from easy, and requires regular updating. Other alternative course structures include the Berkeley Data 8 ``Foundations of Data Science" course (see {\it data8.org}).
\\
\\
Another feature that distinguishes our DSC101 from similar courses is its practical flavor. Apart from its traditional statistics rigor, DSC101 also has a strong industry flavor: it has an emphasis on the practical aspects, and many examples are taken from applications in industry; the idea is to provide students with authentic data experiences (Grimshaw 2015). The first author has previous data science experience in industry, and in designing this course we use examples from data science in industry and carry out some reverse engineering to decide what topics, projects, and other components are to be included so that students can gain experience with the practical demands of industry. For example, we choose to use R as the programming language for this course, due to the increasing popularity of R in industry. Similarly, given that a data scientist typically spends about 60-70\% of their daily work in pre-processing the data, including the collecting, cleaning and transforming of the data, we have a project that requires students to collect and process unstructured auto sales data from the web, and students are encouraged to use {\it Python} for this purpose.  
\\
\\
The remainder of this paper is structured as follows. First we present two examples of data science applications to motivate the concept of the data science life cycle in Section~\ref{section:lifeCycle}. This is followed by a discussion of the theoretical basis for student activity in Section~\ref{section:theoryBasis}. Then we discuss philosophies of the course design in Section~\ref{section:courseOverview}. This is followed by an introduction in Section~\ref{section:lectureTopics} of individual pieces in the data science life cycle, namely, the generation of questions, data collection, various topics in exploratory data analysis, and then linear regression and hypothesis testing. Finally, we conclude with remarks.
\section{The data science process and life cycle}
\label{section:lifeCycle}
\begin{figure}[h]
\centering
\begin{center}
\includegraphics[scale=0.36,angle=0,clip]{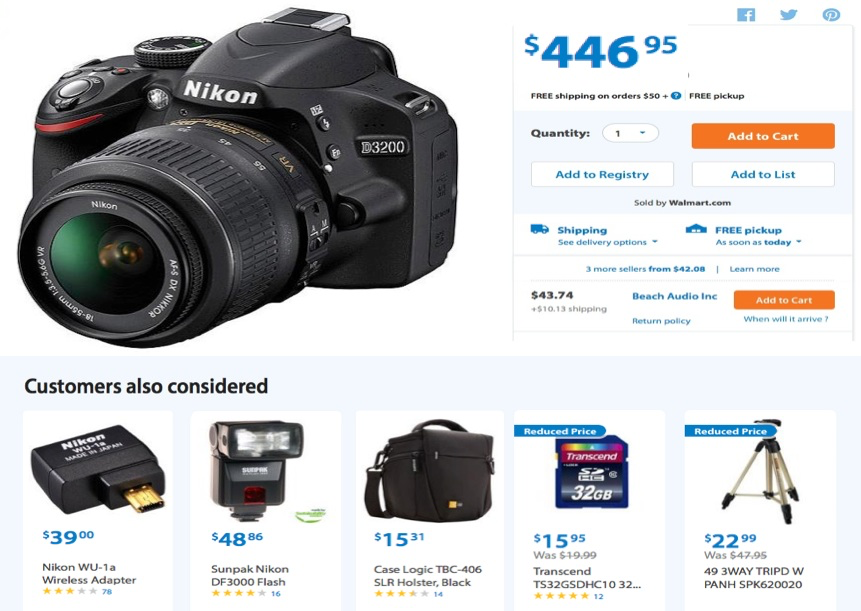}
\end{center}
\caption{\it Items recommended when a camera is clicked. Courtesy {\it walmart.com}.}
\label{figure:cameraBAB}
\end{figure}

As stated in Section~\ref{section:Intro}, our design of DSC101 centers around the data science life cycle. In this section, we will explain the data science life cycle in detail, through two examples. One is about a large-scale study in untangling the relationship among smoking, low birthweight, and infant mortality. The second is on how an e-commerce web site may use historical transaction records to build an item recommendation engine. As will become clear shortly, these represent two different modes of how a data product could be built, and  correspondingly, two different paths in the data science life cycle. 
\begin{figure}[h]
\centering
\begin{center}
\includegraphics[scale=0.34,angle=0,clip]{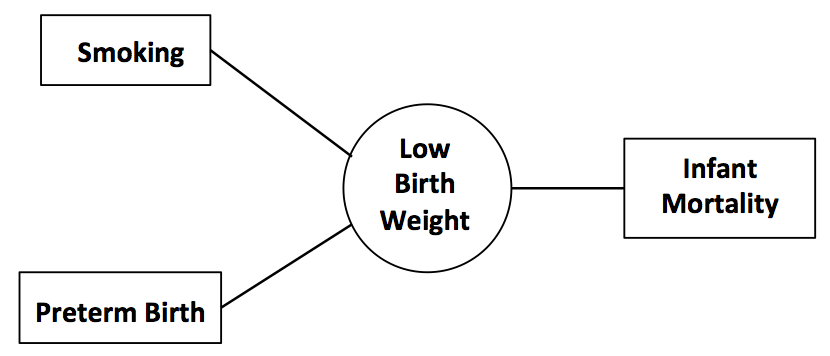}
\end{center}
\caption{\it Smoking, low birthweight, and infant mortality. The link between nodes indicates association instead of causation.}
\label{figure:lbwMortality}
\end{figure}
\\
\\
The first example is from a noted study---the Child Health and Development Studies, carried out by Yerushalmy (1964, 1971) in the 1960's on how a mother's smoking, low birthweight of infants, and infant mortality, are related. Several prior studies, e.g., Simpson (1957), suggested a much greater proportion of lower birthweights (i.e., less than 2,500 grams for newborns in the US) among smoking mothers than nonsmokers. Meanwhile, low birthweight was a strong predictor of infant mortality. Is smoking related to infant mortality? Data were collected for all pregnancies (about 10,000 cases before 1964, and later increased to about 15,000) between 1960 to 1967 among women in the Kaiser Foundation Health Plan in Oakland, California (Nolan and Speed 2000). The data includes the baby's length, weight, and head circumference, the length of pregnancy, whether it is first born or not, age, height, weight, education, and smoking status of the mothers, as well as similar information about the father etc. Yerushalmy's 1964 study confirmed prior claims on a greater proportion of low-weight births but no higher mortality rate for smoking mothers. Yerushalmy collected more data, for about 13,000 pregnancies, and refined his research focus on low birthweight infants. This led to the unexpected finding that, among the low birthweight infants, those from a smoking mother actually survived considerably better than otherwise. A later study (Wilcox 2001), directed by Allen Wilcox on a much larger data set of about 260,000 births in the state of Missouri (1980-1984), resolved the low birthweight paradox and found that infant mortality was primarily caused by other factors, such as preterm birth. Wilcox writes:
\begin{quote}
    ``the mortality difference must be due either to a difference in small pre-term births or to differences in weight-specific mortality that are independent of birthweight. This demonstrates the central importance of pre-term delivery in infant mortality, and the unimportance of birthwieght'' (Wilcox 2001, p. 1239).
\end{quote}
The second example is about item recommendation on an e-commerce web. An e-commerce vendor would typically collect traces of every `mouse click' when a user visits its web, including items a user clicks, views, or purchases. Such data is often called {\it clickstream} data, which contains fairly rich information about users' purchase behavior: for example, the most popular items, items a user typically buys together (called ``co-bought items''), and geographical patterns in users' purchase behavior. Such user behavior profiles can be used to recommend selected items to the user, or to select appropriate contents to show the user when he enters a new page. This is called item recommendation or personalization. 
For example, in Figure~\ref{figure:cameraBAB}, a user has clicked a {\it Nikon} camera. The co-bought statistics from historical data, taking into account of item prices, can be used to decide which items to display that would lead to the most user clicks or the most profit for the vendor. 
\\
\\
The first example describes the path taken by traditional statistical analysis. That is, one starts with a question in mind, then collects data, followed by data analysis, more data, refined question, and then a conclusion. The second example describes an alternative path, where large data have been collected (e.g., as a by-product of normal business operations) but it is not clear what to do, so one will need to come up with a relevant question (such as ``what behaviors predict purchases?'') through some preliminary analysis on the data, and then conduct data analysis until reaching a conclusion or outcome. One thing in common is that both examples consist of the same set of tasks: data collection (including data cleaning and pre-processing), questions, analysis and outcome (a conclusion, a model or data products etc). Data analysis can be either exploratory data analysis (EDA) in which one explores the data and constructs hypotheses, or confirmatory data analysis (CDA) in which one tests prespecified hypotheses via a model on variables of interest. One reaches a conclusion or outcome either by EDA or CDA. Note that some steps may be repeated multiple times. Among these tasks, there is a dependency: some tasks only start upon the completion of those proceeding ones. Each data science application has a {\it start}, followed by a series of tasks, and finishes with an {\it end}. We use a concept called {\it process} to describe this, in analogy with the process concept used in computer operating systems. Interdependent tasks are linked by a (directed) arrow---the task pointed to by the arrow only starts when the task at the source of the arrow completes. Putting these together, we arrive at a directed graph. This is the {\it data science life cycle}, similar to the software life cycle (Langer 2012) used in software engineering. 
\begin{figure}[h]
\centering
\begin{center}
\includegraphics[scale=0.35,angle=0,clip]{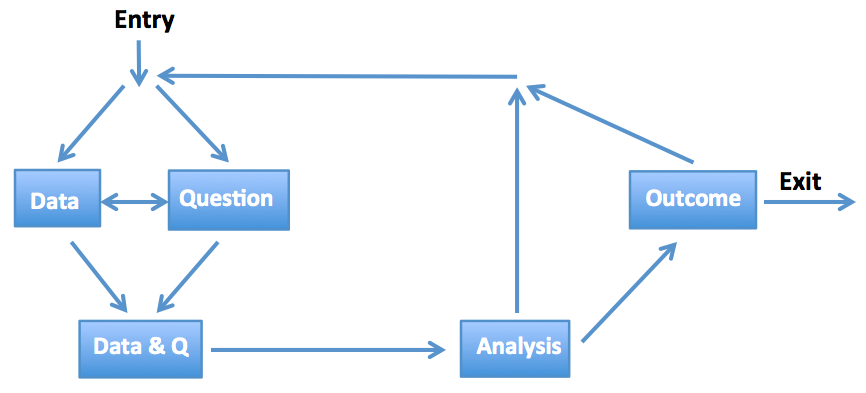}
\end{center}
\caption{\it The data science life cycle.}
\label{figure:dsLifecycle}
\end{figure}
\\
\\
Figure~\ref{figure:dsLifecycle} is our proposed diagram for the data science life cycle. 
``Data \& Q" indicates a state in the data science life cycle such that, at the current state, one has collected the data and formulated a study question.
\begin{figure}[htp]
\centering
\begin{center}
(a)~~~~ \includegraphics[scale=0.29,angle=0,clip]{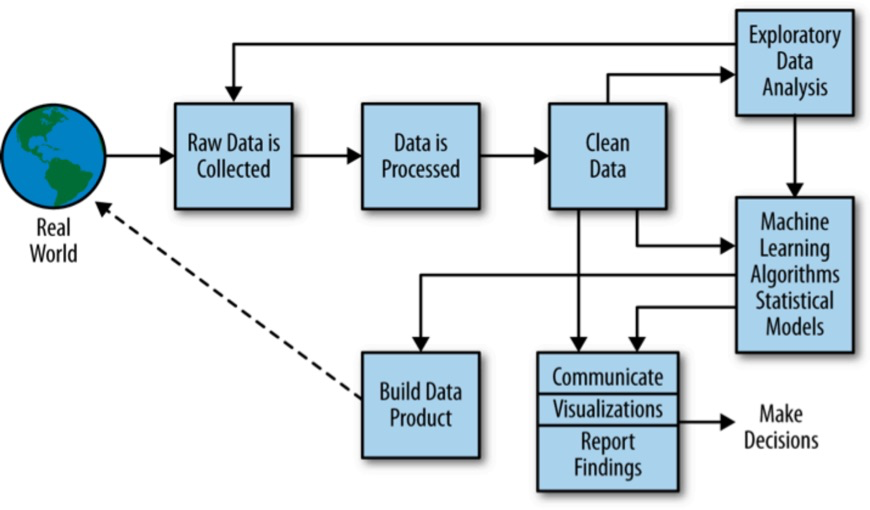}\\[0.08in]
(b)~~~~ \includegraphics[scale=0.41,angle=0,clip]{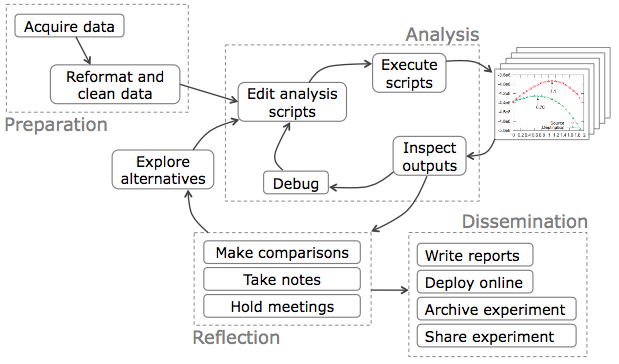}\\
(c)~~~~ \includegraphics[scale=0.24,angle=0,clip]{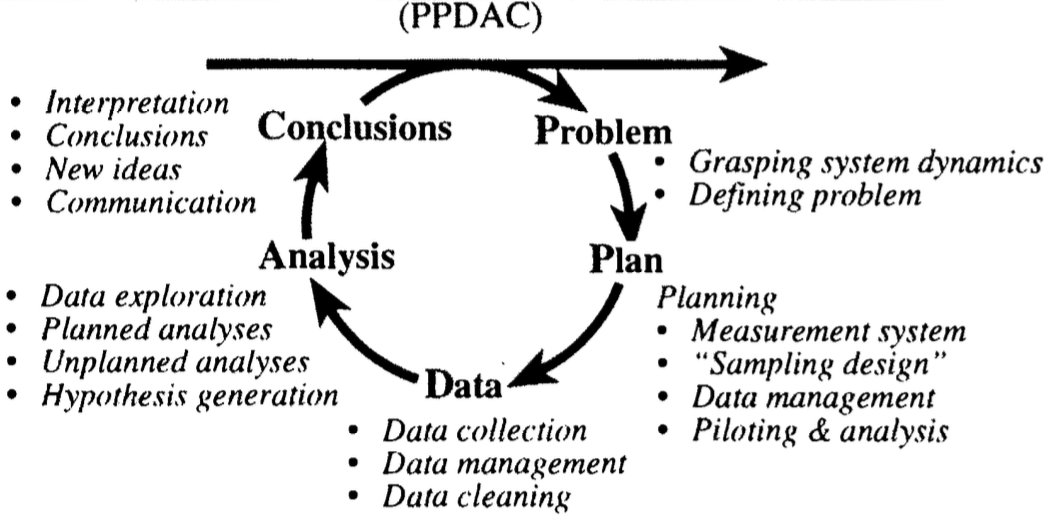}\\[0.05in]
(d)~~~~ \includegraphics[scale=0.76,angle=0,clip]{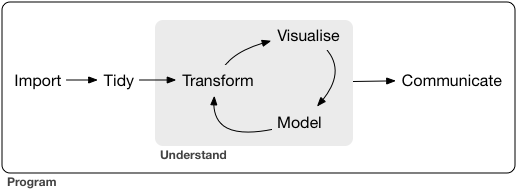}
\end{center}
\abovecaptionskip -3pt
\caption{\it Alternative diagrams related to the analysis of data. a) The data science process diagram of Schutt and O'Neil; b) the data science workflow of Guo;
~c) the PPDAC cycle; ~d) the Wickham-Grolemund data science cycle.
}
\label{figure:dsAlternateCycle1}
\end{figure}
\noindent 
\\
\\
Similar models or diagrams have been proposed for data science during the last few years, for example, Schutt and O'Neill's data science process diagram (O'Neil and Schutt 2013), Phillip Guo's data science workflow (Guo 2012), the PPDSC cycle (Wild and Pfannkuch 1999), and the Wickham-Grolemund data science cycle (Wickham and Grolemund 2016). These are illustrated in Figure~\ref{figure:dsAlternateCycle1}. However, there are major differences from our model. 
Schutt and O'Neil's diagram focuses on the data and describes intermediate stages in the building of a data product, so it is essentially a data cycle. Guo's workflow model describes the dependency of various tasks in a data science project setting; it includes many details and may not be general enough. The PPDAC Cycle is the closest to ours in the sense that it consists of one possible paths in our diagram. The Wickham-Grolemund data science cycle takes a data-centered approach and form a cycle by including various operations to the data. Our model focuses on the \textit{tasks} in data science, and allows the interaction between tasks and their repetitions, as well as the possibility of having a clearly defined question in mind at the start. 
\section{Theoretical basis for student activity}
\label{section:theoryBasis}
\begin{figure}[h]
\centering
\includegraphics[scale=0.52,angle=0,clip]{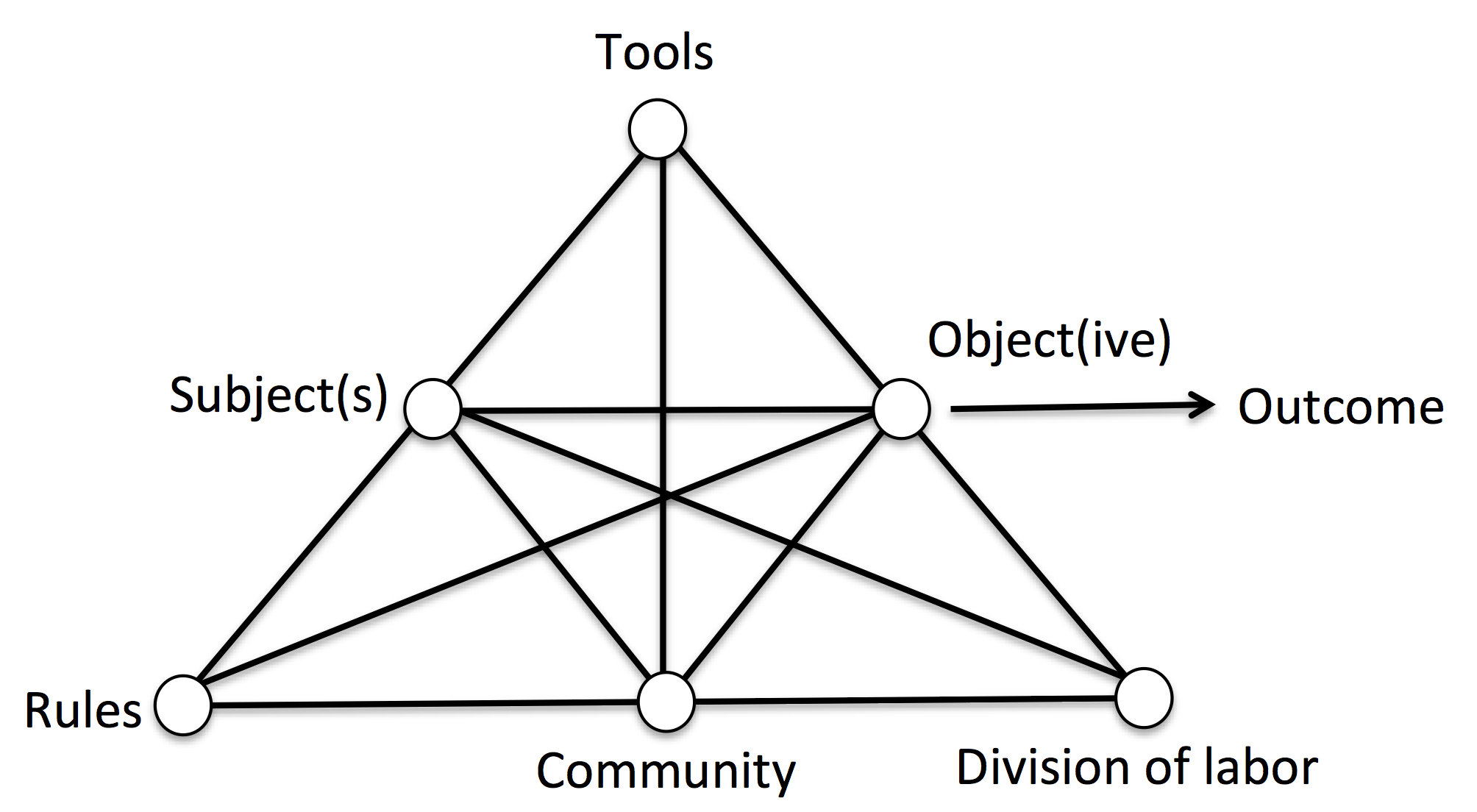}
\caption{\it A generic activity triangle}
\label{figure:GenericActivityCycle}
\end{figure}

Semester-long undergraduate courses are designed with specific aims and learning outcomes in mind, and college or university administrators require these to be explicitly articulated. Additionally, instructors bring with them a theoretical stance on how a course will be executed over a semester. This theoretical aspect to course design and implementation is rarely explicitly articulated (and sometimes  not even by the instructor to themselves!) A clearly articulated theoretical basis for the design of an introductory data science course is of great importance because it sets the scene for how a course unfolds throughout a semester. Of the many differing educational theories that might productively apply to designing data science courses, \textit{activity theory} (Leontiev 1978; Davydov et al. 1982; Raeithel 1991) provides a coherent and productive foundation, and we discuss how aspects of activity theory interact with the data science life cycle. 
\\
\\
As this is only a first course in data science, and usually offered at the beginning of the first year, students typically have not acquired a strong background in calculus or statistics, so we need to begin by working with what intellectual tools they do have and then introducing them to new analytical and computational tools. We focus in the beginning, mainly on exploratory data analysis, and concepts related to various parts in the data science life cycle. This includes introduction to concepts or tools such as sampling, descriptive and summary statistics, data visualization and graphical tools, moving on progressively to the use of such tools as principal component analysis, clustering, linear regression and hypothesis testing. 
\\
\\
A major point is the following: the tools are introduced in order to enable students to effectively transform raw data into something more useful. The focus is on the raw data, the motivation to transform them---the \textit{objective}---and the tools used to effect those transformations. This is the opposite of a scenario in which techniques of data analysis are taught with artificially designed and relatively simple toy data (that is, students practice tool use in the absence of appropriate or realistic data). Becoming a useful and skillful data scientist requires addressing {\it the full complexity of data}, and finding appropriate tools to effect insightful transformations on those data. This is the central reason why activity theory drives so much of our thinking in course design for data science: from an activity theory perspective the \textit{context} of data science for these beginning students is the raw data, questions posed about those data, agreed objectives, and transformation of the data by activity, utilizing analytical tools, in a cyclic process. By this perspective, introductory data science is contextualized for the students as a meaningful, empowering process. In many academic courses students do exercises and practice on toy data sets to complete homework exercises and study for an examination in order to get a satisfactory grade. The reality of the context makes the DSC101 course quite different from this.
\\
\\
Specific curriculum instances of activity theory are often described in terms of an ``activity triangle'' (see, for example, Engestrom 1991, 1999, 2000 and Price et al 2010). Typically, these activity triangles have a structure as illustrated in Figure~\ref{figure:GenericActivityCycle}. 
\noindent
\\
\\
An activity triangle encapsulates the interrelationship between the main constituents of a curriculum activity as conceived by activity theory. 
As a specific example consider a traffic data example (see more details in Section 7.1), which consists of the starting point and destination of each trip and time stamp at each road during the trip.  The activity starts with raw material which is real traffic data.
\\
\\
The \textit{community} is the class of students and the instructor, but may also include an audience, other than the instructor, for whom the students are to build a data product such as a predictive model, and write a report. For example, the traffic data may well have come from someone who wants to know certain things about the data, so in this case students write reports for that person, who is also part of the community 
\\
\\
The \textit{subject or subjects} consist of an individual student or small groups of students working together to produce an outcome, typically a written report of an analysis or a data product. 
\\
\\
The \textit{tools} are usually the software tools, such as the R programming language and conceptual tools, such as regression or clustering techniques, that students can bring to bear on the objective. 
\\
\\
The \textit{object, or objective}, is determined through discussion by students, the instructor and any external client, and in the case of the traffic data example may involve such things as determining traffic bottlenecks at particular times of day. 
\\
\\
The \textit{rules} vary from activity to activity, and may include such general things as avoidance of plagiarism, appropriate referencing of sources, cooperation within and between student teams, sharing of findings, ethical behavior, and responsibility for meeting deadlines. 
\\
\\
\textit{Division of labor} can work in several ways including different students within a team taking charge of different aspects of analysis, or different teams focusing on different aspects of an objective with the aim of pooling findings. 
\\
\\
The same data set may---and usually does---generate a number of different activities and objectives as students ask further questions about the data, and set out to examine their determined objectives. When this happens the objective will change, the subjects may change in that students may form new groups, spontaneously or at the instructor'€™s direction, the division of labor may change, and the tools will most likely need to be modified and new tools brought to bear on achieving the objective.
\\
\\
An activity triangle, as realized in a specific curriculum module, is coordinated with the data science life cycle. Although many variations are possible from activity to activity, as described above, it is common that certain aspects of an activity triangle stay fixed throughout a semester: typically the \textit{subjects} are the students; the \textit{community} is the class of students and the instructor; the \textit{rules} are articulated at the beginning of semester and stay more or less fixed; and commonly the \textit{division of labor}, either within or between groups, stays much the same. The data science life cycle impacts the activity triangle, and vice versa, from question to objective, analysis to tools, and outcome to conclusion.
\\
\\
As students are engaged in a specific project---some of which are detailed below---and cycle through the data science life cycle, a new activity triangle emerges in which new questions inform new objectives, new analyses require new tools, and new outcomes provide new conclusions. Thus one sees a dynamic sequence of activity triangles as progress on a project involves cycling through the data science life cycle. The activity triangles inform the data science life cycle in that they describe how the various aspects of the data science life cycle are implemented through activity.
\\
\\
We focus on the practice and craft of data science---part of what it means to become apprenticed as a beginning data scientist. This does not mean, however, that something akin to Lave's situated action model (Lave 1988, Lave and Wenger 1991), in which one learns by self-directed, novice participation in a communal activity, provides a better theoretical model for designing a data science course than does activity theory. The essential feature of activity theory that is helpful in this regard is that an object comes \textit{before} an activity based on that object, and motivates the activity (Nardi 1996). While learning to become a data scientist through behaving as if one were an apprentice, thrown into an ongoing field of activity, can be a positive and highly educative experience---and is the motivation for many student internships---our focus in beginning data science courses is on activity motivated by student desire to transform data, using tools they have at hand, or are capable of developing: this constitutes an ``object'' (or ``objective'') in activity theory. Data is transformed through activity that relates to an objective usually coming from a naturally arising question about the data. 
\section{Course design}
\label{section:courseOverview}
Activity theory helps us focus on two aspects of DSC101 that are important to its success. The first is that the data is real-world data for which a question---sometimes rather vague---is naturally proposed. For example (see also Section~\ref{section:askQs}), given a collection of traffic data for many trips, including starting and end point as well as timestamp at each road during the trip, what questions could students ask that have the potential of becoming a data product? This aspect of the DSC101 course is important in focusing students on an end-product of their studies in data science: a rewarding and satisfying career. From the beginning, students in DSC101 gain a lived experience of what constitutes both the practical and conceptual aspects of the working life of a data scientist. The second aspect of DSC101 highlighted by an activity theory perspective is \textit{empowerment}: the extent to which the activities and tools used in those activities actually empower students to do something satisfying. Students in DSC101 should never complain: ``When will we ever use this?'' The answer is obvious from the nature of their activity in attempting to answer questions about real-world data with tools provided to them, or built by them. 
\\
\\
The design of our DSC101 course centers around the data science life cycle and the activities that involve. The course starts with an introductory lecture of data science with two goals in mind. One is to give students a sense that data entails value, another that it is possible to make a difference, to influence outcomes, by leveraging values from the data. We introduce numerous interesting stories from a variety of fields, ranging from science, finance, metrology, sports, to Internet and e-commerce, on how  insights can be obtained from the data through models and analytical tools. Of course, these stories also convey an idea to students of what constitutes data science, and how their activity, on raw data, with specific objectives, can transform that raw data to insightful outcomes through the use of appropriate tools. Then the data science life cycle is introduced, followed by various parts of the cycle, including asking interesting questions from data, data collection, exploratory data analysis, modeling, and confirmatory data analysis. 
\\
\\
To reflect the practical aspect of this course, also due to its growing popularity in the data science community, we dedicate two weeks of lectures for R programming (Verzani 2008), which is the programming language used for instruction and student projects. There are many alternatives to R, 
the free and open source nature of R together with a very large and diverse R user community make a relatively compelling case for including R as a basic programming language and data analysis tool. Through being inducted into the R ecosystem students are exposed to a huge network of open data analytic resources and tools by learning the basics of R programming: it's not simply a useful and widely used tool they learn---it's also a huge and diverse community of potential support. People who use R, write R packages, and provide instruction in, and support for, R come from a widely diverse collection of backgrounds, so exposing beginning data science students to a vision of data science that cuts across numerous disciplines.
\section{Topics covered in the course}
\label{section:lectureTopics}
As described earlier, topics covered in our DSC101 are individual parts in our data science life cycle. In particular, Section~\ref{section:askQs} corresponds to ``question", Section~\ref{section:askQs} to ``data", Section~\ref{section:EDA}, Section~\ref{section:lm}, and Section~\ref{section:CDA} corresponds to ``analysis" part of the data science life cycle, respectively. In this section, we will describe each
of the topics in detail.  
\subsection{Asking interesting questions}
\label{section:askQs}
Asking informed questions, from data or given evidence, is one of the most crucial parts of traditional sciences:  it forms the start of a scientific investigation. On the other hand, it is one of the primary driving forces behind the recent explosive growth in data science applications. Imagine that an e-commerce vendor has collected huge user access data; what new business models can it generate? If a search engine has collected a large collection of searched keywords, how could such data be utilized? It is possible to use such data to optimize the selection of advertisements and their placement in a page, or even to improve the design of the search engine.
\\
\\
To paraphrase Brown and Keeley (2007, p. 3) in the context of a DSC101 course: Questions about data require the person asking the question to act in response. By our questions, we are saying: I am curious; I want to know more. The questions exist to inform and provide direction---an objective---for all who hear them. The point of questions is that one needs help and focus in obtaining a deeper understanding and appreciation of what might be in the data.
To inspire students to think and appreciate the value of data, and ask good questions, students are encouraged to ask questions for any data to which they may have access. As an example, in-class groups are formed among students to discuss potentially what one could do with large traffic data. 
\begin{quote}
{\it Suppose one is given traffic data of a city. Data includes about 30 million records of vehicles with each consisting of:  the starting point and destination of each trip, and time stamp at each road during the trip. The same car may have multiple entries in the records. There are two cases: knowing or not knowing the auto plate. 
What can one do with such data?}
\end{quote}
\subsection{Details of R programming}
R is chosen as the programming language for the course, recognizing the growing importance of R programming in data science as well as its great utility in modeling (modeling is offered as a senior level undergraduate data science course at the University of Massachusetts Dartmouth). Topics covered include three parts. 
\begin{itemize}
\item The first is on programming language features. This includes data structures such as lists, vectors, arrays and matrices, data frames etc; structured programming constructs such as loops, conditional statements and functions etc; data and text manipulation (including regular expressions) tools, file I/Os (including excel spread sheets) etc. 
\item The second is on the statistical aspect of R, which covers R functions to generate data of various distributions, and R functions for statistical tests etc. 
\item The third is on R functions for graphics and visualization. As an elementary course in data science, only R functions or simple graphical tools related to basic plotting functionalities are discussed. 
\end{itemize}
To sharpen the programming skills of the students, very simple algorithms related to searching and text manipulation are introduced. Programming exercises are assigned as labs, and programming questions, such as analyzing the program output and implementing a simple function, are included in the exams. Sample R code is provided for most of the examples, so that students can try R programming on their own and gain hands-on experience. 
\subsection{Sampling and data collection}
\label{section:dataCollect}
Data collection is an important aspect of data science. In DSC101, the idea of random sampling and sampling techniques such as simple random sampling and stratified sampling are introduced. To better appreciate the idea of random sampling, several types of mis-uses of sampling are discussed, including {\it sampling from the wrong population, convenience sampling, judgement sampling, data cherry-picking, self-selection, and anecdotal examples}. Each of these is discussed with a story, selected from the news or from the instructor's experience. Before a formal analysis of each story, time is allocated for students to think and to form group discussions to see if there is anything potentially wrong in the story. Students are also encouraged to share their own examples. As a practice, students are assigned a lab to collect auto sales data, including sales prices and the age of their favorite car model, and judge if their data collection suffers from any sampling bias. Such learning by doing practice may improve students' interest in the course.
\subsection{Exploratory data analysis}
\label{section:EDA}
Exploratory data analysis (EDA) was pioneered by J. W. Tukey in the 1960's (Tukey 1977). It refers to various things one would try out before a formal and often complicated data analysis, and is therefore often viewed as a preliminary data analysis. It is typically applied in situations when one wishes to know more about the application domain, and EDA often helps one gain a better sense of what the data looks like, which may be suggestive in the choice of a model or data transformation. 
Similarly when one has  data but does not have a well-defined question, exploring the data to discover patterns or regularities may inspire interesting questions. Of course, sometimes EDA may be sufficient if the question of interest is rather simple or the underlying pattern is salient enough. Common tasks in EDA include  the following: descriptive and summary statistics, graphical visualization, data transformations, clustering etc. We discuss each of these in the following.
\subsubsection{Descriptive and summary statistics}
Descriptive and summary statistics are very helpful in data analysis. From such statistics, one can often get a ball-park idea of the data distribution. These are also useful in presenting data or communicating  results to other people, especially when graphical visualization is not possible. Three types of descriptive or summary statistics are introduced in DSC101. The first is for the measure of location in the distribution, including {\it mean, median, mode}, and the more general {\it quantiles} and {\it percentiles}. The second is for measures of dispersion, including {\it variance} and {\it standard deviation}. The third is about the shape of the data distribution. This includes a measure of asymmetry of the data, {\it skewness}, and a measure of the peakedness of the data, {\it kurtosis}. 
\subsubsection{Graphics and data visualization}
Data visualization is an important part of EDA, and also a useful tool for communicating results. It is being used more and more in the practice of data science, for example one may see plots or charts in almost every issue of the \textit{New York Times}, and the \textit{Guardian} newspaper, in its various country and international editions. 
\\
\\
This part of the course starts with guidelines, or rules of thumb, for a useful visualization. Note that our focus is the visualization of data instead of abstract concepts (Yan and Davis 2018); here one seeks to understand the data or information behind by displaying aspects of the data. Then a collection of graphical tools are introduced, including basic tools such as bar, pie, Pareto charts and their stacked or grouped version; statistical graph tools such as histograms, box plots, stem-and-leaf plots; as well as tools suitable for the visualization of multivariate data.
Some interesting data sets are used in introducing the graphical tools, for example the US crime and arrest data, the US statewide mean January temperature for a given year and the mean during the last century. Students use the tools and example R code to visualize the data, then share what they observe from the graphs or other visualizations, and give interpretations. 
To better appreciate the effect of graphical visualization (Nolan and Perrett 2016), in-class discussions are formed where students are given a data set, such as a multiway contingency table, and then tasked to design their own way of visualization, and designs from different groups are compared. This is a good opportunity for students to apply what they learn with creativity, and greatly motivates students' interests in the course. Indeed quite a few students view this as the best part of the course. 
\\
\\
For the visualization of multivariate data, tools such as bubble plots, Chernoff faces (Chernoff 1973), and radial plots are introduced. In particular, students find Chernoff faces interesting and intuitive, and that helps them to gain insights: for example on the US crime or political ideology by states. Principal component analysis is another tool introduced to visualize multivariate data and for dimension reduction. 
\subsubsection{Data transformation and feature engineering}
Feature engineering refers to the creation of new features from the data, or, combining or transforming existing features into new ones that suitably represent or reveal interesting structures or patterns in the data. It is a task on which data scientists typically dedicate major time. It is crucial to the success of many modeling tasks. Better features often lead to better results, more flexibility, and better interpretation of the results. While the entire world has been excited about the success by an emerging machine learning paradigm, deep learning (Hinton and Salakhutdinov 2006, LeCun 2015), on the automatic discovery of useful features from  data, applications beyond image, speech, and natural language processing still rely heavily on feature engineering. As students in DSC101 are unlikely to have any prior data science experience, we only introduce the concept of feature engineering and focus on the easiest part---data transformation. Data transformation is needed when different features have drastically different numerical scales, or when the underlying pattern or regularity in the data becomes more salient due to data transformation. Topics discussed include Tukey's idea of ``straightening the plot" (an idea that guides data transformation from human perception) (Tukey 1977), and the Box-Cox power transformation (Box and Cox 1964). Several transformations frequently used in practice are discussed. This includes logarithmic or square root transformation, data standardization to mean 0 and variance 1, linear scaling of the data to a range $[a,b]$, non-linear bucketing of the data (e.g., assign a numerical value 1 to income lower than 20,000, and 2 for the range [20,000, 50,000) and so on).    
\subsubsection{Clustering}
In practice, data are often heterogeneous. This is due possibly to spatial, temporal effects, or differences in other characteristics (e.g., male or females often have very different life style or shopping behavior). Heterogeneity is especially common for big data. It is often desirable to divide the data so that data in the same subgroup is of a similar nature. One way to achieve this is via clustering. Three classical clustering algorithms are introduced, including hierarchical, agglomerative, and K-means clustering (Aggarwal and Reddy 2013). The idea of the algorithms and important properties are discussed. More advanced and modern clustering methods such as model-based clustering (Fraley and Raftery 2002), spectral clustering (von Luxburg 2007), cluster ensemble (Strehl and Ghosh 2003, Yan et al. 2013) etc are not discussed in lecture but may be used for course projects for students with adequate preparation in calculus and linear algebra. 
\subsection{Simple modeling with linear regression}
\label{section:lm}
Simple linear regression is introduced both as a continuation of visualization, in the sense that the regression line is the line that is `close' to most of the data points, and also as a way to summarize data with a simple function. This leads to the concept of modeling. Example models are given that students are likely to have learned in their high school texts or from other courses. For a better appreciation of the concept, students are asked to give their own examples of models. Formulation of simple linear regression is introduced as a least square optimization problem, as well as the concept of $R^2$ as an indicator of the amount of variance explained in the model. The term {\it regression} was discussed, using classical father-son height data. Simple linear regression was naturally extended to multiple regression, using the auto mileage per gallon (MPG) data from the UC Irvine Machine Learning Repository. Before discussing this example, students are asked to make a guess on which factors are important to the gas mileage of a car; after seeing the regression analysis results students would better appreciate the value of data analysis. Relevant R functions for linear regression are introduced, along with discussion of how to read the regression output. Depending on the preparation of students, it may be possible to 
extend the discussion to multiple linear regression as recommended by the revised GAISE College Report (2016). 
\subsection{Confirmatory data analysis and hypothesis testing}
\label{section:CDA}
In the confirmatory data analysis part of DSC101, the statistical framework of hypothesis testing is introduced. There have been lots of controversies on the usage of p-values in recent years (see, e.g., Cumming 2013). However, it is still widely used in industry. For example, many vendors in  industry use A/B testing\footnote{A/B test is the application of hypothesis testing to compare the effectiveness of two alternatives (one termed as ``A" and the other ``B"). It is used widely in industry to compare alternative models or strategies.} and p-values for the comparison of alternative models or strategies. The concept of hypothesis testing is often challenging to students, as it represents a different way of reasoning compared to logic deduction, with which they are likely more familiar. To help students, two analogies are introduced and analyzed, one being the court trial and the other proof by contradiction. This greatly helps students in understanding. An example from  industry is used to explain why hypothesis testing is useful, e.g., A/B test in deciding if a new strategy or model does better than the existing one via hypothesis testing. Several students expressed a view that they liked this part of the course as it seems surprisingly useful for many real world problems.
\subsection{Difference from a statistics course at similar level}
As can be seen, a big part of the course would overlap with a typical statistics course at the similar level. We attribute 
this to the intimate relationship between data science and statistics; we would not expect a data science course to be 
very different from a statistics course. That said, compared to related statistics courses at institutes with which the authors 
are familiar (there is not a similar statistics course at our institute), there are several major differences apart from 
topics apparently missing in these statistics courses (i.e., topics on biases in sampling, feature engineering, visualization 
of multivariate data, PCA, clustering). Similar statistics courses would not be structured by the (data science) life cycle,
and the main theme of the courses here is on leveraging data for insights, conclusions, models, or data products. In a 
similar statistics course, there would not be any motivating lectures on leveraging value from the data, nor is there any 
discussion of data science life cycle in the form of carefully chosen examples or in-class discussions. There would not 
be so much discussion on visualization in a typical statistics course. Also likely 
the data for projects are given instead of asking students to find or scrape data by themselves. Potentially, 
there may also be differences in the execution even if the schedules might look similar. For example, we use many examples 
from the industry (including some from the author's past work), which may not be the case for a typical statistics course.
\section{Other course components}
Section \ref{section:lectureTopics} discusses topics for lectures, yet there are other components of the course not touched, namely, labs or course projects, and presentations. We will briefly discuss these here; for more details, we refer the reader to the sample syllabus in the appendix.  
\subsection{Labs and course projects}
An important part of a data science course is projects. As DSC101 is offered mostly to first-year students, and students typically do not have prior exposure to any programming language, the course project is in the form of several small labs. Each lab touches a major topic in the course, and students are typically given two weeks time to work on each project. Students will write a lab report describing the project, where and how the data are collected, a description of the data analysis procedure, and conclusion, if any. R code is required to submit with the lab report. This is a critically important part of DSC101 because it introduces students to an essential characteristic of a data science professional: the ability to clearly communicate the results of data analysis (see, e.g., Sisto 2009, O'Neil and Schutt 2013).   
\\
\\
The first project is mainly on data collection. Students are required to find data online or from other sources, and then conduct some simple exploratory analysis. One is to download and extract auto sales price for a particular car model from a popular auto sales web, {\it cars.com}, for cars of different years. The average prices are calculated for cars of the same years, and then a price-year plot is produced. The second example is from {\it kaggle.com},
which consists of historical records of airplane crashes since 1908. Students download and process the data, then visualize airplane crashes by year, airlines, and aircraft models. In terms of empowerment, some students became very excited about the notion of data analysis for insights, and started analyzing data related to their own interests. For example, one student chose to analyze data on basketball games, and observed the rising of 3-point shots in recent years; he also made interesting predictions on the strategy of future basketball games. 
\\
\\
The second project is to read an article of data analysis. One example is about analysis on the swimming competitions in the {\it Rio Olympics}. Two interesting phenomena were observed, namely, the noted difference in time between back and forth laps, and the observed disadvantage towards athletes assigned to lower-numbered lanes. Students are required to write a report on how the author uses the data and carries out his analysis to reach his conclusions. Students were asked if there are any biases in the way the author was designing the study. The second part of the project is to have students find two examples of misuse of sampling techniques in collecting data, from recent news or articles.
\\
\\
The third project is about descriptive statistics and sampling techniques. Several data sets are given and students are asked to compute the skewness and kurtosis. The second
part is about sampling techniques, to compare simple random sampling (SRS) and stratified sampling. Students find or generate their own data set that is `heterogeneous', and then compare SRS and stratified sampling on the variation in the sample means if they are to repeat the sampling 100 times. 
\\
\\
The fourth project is the visualization of US population by states, for Census 2000 and 2010, respectively. In particular, students are required to produce an appropriate heatmap on the US map, and then plot a bubble plot on the rate of change in population on the map. 
\\
\\
The last project is about the application of different clustering methods, including K-means, agglomerative, and divisive clustering. Students produce dendrograms and compare the results.
For this project, students are required to do a short presentation for the project of a 10 minutes duration, including questions and answers. As stated above, an important part of a data scientist's job is to communicate a problem of interest, or to present analyses, to other people. We make presentation of projects, and in-class discussion, in addition to written reports an important part of the course. 
\subsection{Assessment}
The students' performance in the course is assessed in all course components, including quizzes, labs, in-class discussion, a midterm, a final exam. Also there are two in-class practice sessions. The idea is to ensure  students go through the relevant course materials and apply these to problem solving. The instructor can observe students performance and provide help on any potential issues students may have. This is allocated to two key topics of the course,  R programming and hypothesis testing. 
The grade breakdown in a typical semester is as follows: quizzes--10\%; in class discussion, practice or presentation--20\%; labs--20\%; midterm--20\%; final--30\%.
Team-based learning is incorporated in in-class discussion or presentation, or labs (students can choose to do it individually, or as a team). 
\subsection{The students, engagement and feedback}
We have been teaching this course since Fall 2015 (this course is offered every Fall). Typically about 40\% of the students are data science majors, with others from a very diverse list of majors, such as mathematics, computer science, biology, electrical engineering, mechanical engineering, accounting, management information systems (MIS) etc. This is not a service course.
\\
\\
We do not offer a similar introductory statistics course at University of Mass Dartmouth. At one other institute, one author taught a similar course, {\it Elementary Statistics}. In DSC101, the students are more engaged. We attribute that to the following based on our observations and feedbacks from students. This course is better motivated with many realistic applications.
The course requires more hands-on from students, for example, students need to try out simple examples using R programming during class. The in-class discussions use topics students are familiar with and that they could apply their creativity. Finally, students have more freedom in choosing their projects using real data.
\\
\\
Feedback from students suggests that they generally like the in-class discussion, the hands-on exercise on examples discussed in class, the exam problem on data visualization, and also the freedom in choosing problems for their projects. 
\section{Conclusion}
\label{section:conclusion}
We have briefly introduced a first course in data science offered at the University of Massachusetts Dartmouth since Fall 2015. To facilitate our discussion, we clarified 
our viewpoints on what data science is, and  introduced the notion that data entails value yet to be explored. Our design of the course is both principled and practical. The design centers around the data science life cycle---topics covered 
in the course correspond roughly to individual pieces in the life cycle. That is, data collection, the generation of a study question, data analysis, how to draw conclusions, and how to communicate results. As a first course in data science, our focus is on the 
motivation and concepts, and the formal analysis part is limited to exploratory data analysis, linear regression and hypothesis testing. The practical aspect of the course is reflected in several ways. Our design of the course has incorporated many elements from current data science practice. We use the popular R programming language for instruction, students hands-on exercises, and projects (but we also encourage the use of Python for projects). Our examples and the data used for course projects are mostly from real world applications. In terms of empowerment, the course has been fairly successful in that at the conclusion of this course, students can comfortably carry out elementary data analysis using R and tools introduced in the class, on varied realistic, and real, data sets. Some students even started analyzing datasets related to their own interests, for example, the basketball/baseball games data, the {\it Zillow.com} real estate data, etc. One thing worth noting is that this course has managed to attract several students from other majors to our data science program. We hope that our DSC101 course can benefit educators who are new in the field, or students who are interested in data science. 
\section{Appendix} 
\label{section:appendix}
\subsection{A sample weekly schedule of DSC101}
\label{section:sampleSchedule}
A sample weekly schedule of DSC101 can be seen in Table~\ref{table:sampleSchedule}. The class meets twice a week for a 75 minute session.
\begin{table}[ht]
\begin{center}
\begin{tabular}{c|l}
\hline
Week           &  Topics\\
\hline
1           &  {\it Introduction to data science}\\
		&  {\it The data science life cycle}\\
2-3           & {\it R programming} \\
4     & {\it Concept of sampling and potential bias}\\
5           & {\it Simple random and stratified sampling} \\
		&  {\it Descriptive and summary statistics}\\
6     & {\it Data visualization (principle and basics)}\\
7    & {\it Data visualization (statistics)}\\
 &  {\it Data visualization (bubbles, maps etc)}\\
8     & {\it Data transformations and feature engineering}\\
	& {\it Midterm}\\
9     & {\it Visualization of multivariate data }\\
	 &  {\it Principle component analysis }\\
10      & {\it Agglomerative and divisive clustering }\\
	& {\it K-Means clustering}\\
11            &  {\it Concept of modeling} \\
		&  {\it Simple linear regression} \\
12          & {\it Multiple regression}\\
		& {\it Introduction to hypothesis testing}      \\
13         &{\it t-test}\\
            & {\it Two-sample and A/B test}                  \\
14         &{\it In-class practice of hypothesis testing problems}\\
            & {\it Project presentation}                  \\
15        & {\it Final exam}\\    
\hline
\end{tabular}
\caption{A sample weekly schedule of topics covered in DSC101.} 
\label{table:sampleSchedule}
\end{center}
\end{table}
\noindent
\\
\\
This weekly schedule was designed by statistics faculty. If a computer science faculty were to teach such a course, they could still use the data science life cycle to structure the course. They could replace several parts of the course (e.g., those with a statistical flavor) with a computer science flavor, and possibly focus more on the implementation aspects of data science. For example, they could teach Python instead of R, given the fact that Python is used more for tasks such as the
processing of texts and unstructured data (both R and Python are popular programming languages in data science practice). They could structure the data visualization part with the implementation of visualization and visual analytics from a human-computer interaction (HCI) perspective. They could replace topics such as PCA with data mining topics such
as association analysis, or frequent itemsets mining. For data collection, they might focus more on practical sampling algorithm (possibly in a big data setting), or tools from Python for
data scraping, for example.
\section*{Acknowledgements}
We thank the editors, the associate editors, and anonymous reviewers for their helpful comments and suggestions.


\end{document}